\documentclass[12pt]{article}
\usepackage{euscript}
\usepackage{amssymb}
\usepackage{amsfonts}
\usepackage{amsbsy}
\usepackage{amsmath}
\usepackage{epsfig}
\usepackage{amsthm}
\usepackage{amscd}
\usepackage{amstext}

\usepackage[pdftex]{hyperref}

\def\ben{\begin{equation}}
\def\een{\end{equation}}
\def\half{{\textstyle{\frac{1}{2}}}}

\let\a=\alpha  \let\g=\gamma  
   \let\k=\kappa
  \let\n=\nu   \let\r=v

\let\w=\omega \let\G=\Gamma

\let\pa=\partial
\def\be{\begin{equation}}
\def\ee{\end{equation}}
\def\ba{\begin{array}}
\def\ea{\end{array}}

\def\dalemb#1#2{{\vbox{\hrule height .#2pt
       \hbox{\vrule width.#2pt height#1pt \kern#1pt
               \vrule width.#2pt}
       \hrule height.#2pt}}}

\newcommand{\bea}{\begin{eqnarray}}
\newcommand{\eea}{\end{eqnarray}}
\newcommand{\tr}{{\rm tr} }

\def\R{{{\mathbb{R}}}}

\def\ocal{{\mathcal{O}}}

\begin{document}

\begin{center}

{ \large {\bf Generalized Lifshitz-Kosevich scaling at quantum criticality  \\
from the holographic correspondence}}

\vspace{1.2cm}

Sean A. Hartnoll and Diego M. Hofman

\vspace{0.9cm}

{\it Department of Physics, Harvard University,
\\
Cambridge, MA 02138, USA \\}

\vspace{0.5cm}

{\tt  hartnoll@physics.harvard.edu, dhofman@physics.harvard.edu} \\

\vspace{1.3cm}

\end{center}

\begin{abstract}

We characterize quantum oscillations in the magnetic susceptibility of a quantum critical non-Fermi liquid. The computation is performed in a strongly interacting regime using the nonperturbative holographic correspondence. The temperature dependence of the amplitude of the oscillations is shown to depend on a critical exponent $\nu$. For general $\nu$ the temperature scaling is distinct from the textbook Lifshitz-Kosevich formula. At the `marginal' value $\nu = \half$, the Lifshitz-Kosevich formula is recovered despite strong interactions. As a by-product of our analysis we present a formalism for computing the amplitude of quantum oscillations for general fermionic theories very efficiently.

\end{abstract}

\pagebreak
\setcounter{page}{1}

\section{Results and background}

A central question in the theoretical characterisation of non-Fermi liquids is the fate of the Fermi surface. For instance the `strange metal', and perhaps quantum critical, regions of the cuprate or heavy fermion phase diagrams separate phases with very distinct energy-momentum distributions of fermions. This is seen in many experimental probes, a recent discussion with strong overlap with the concerns of the present paper is \cite{taillefer}.

Quantum oscillations are a robust feature of systems with a Fermi surface \cite{abrikosov}. The recent and ongoing experimental observation of quantum oscillations in the copper oxide high temperature superconductors \cite{Nat1, Nat2, Nat3, PRL1, PRL2, PRL3, PRL4} is reinvigorating theoretical approaches to the subject (e.g. \cite{senthil, sachdev2}). Present measurements are, perhaps surprisingly, consistent with textbook results for quantum oscillations in Fermi liquids. However, as new regimes are investigated, in these and other quintessentially non-Fermi liquid materials, it will be crucial to have theoretical templates available for comparison. For instance, the exciting recent results of \cite{suchitra} show that the effective quasiparticle mass, as read off by fitting quantum oscillations to the Fermi liquid formula, appears to diverge as one approaches a metal-insulator quantum phase transition. A similar divergence is observed in heavy fermion compounds \cite{heavy}. Here one encounters a theoretical hurdle; the most interesting regimes are often strongly coupled, and perturbative quantum field theory treatments may not fully capture the physics of interest. Recent work investigating the effect of interactions on quantum oscillations includes \cite{manybody, marginala, stamp, subir}. In this paper we will use the inherently non-perturbative `holographic correspondence' (see e.g. \cite{McGreevy:2009xe, Hartnoll:2009qx} for relevant introductions) to give a controlled computation of quantum oscillations in the magnetic susceptibility of a strongly interacting quantum critical non-Fermi liquid. We will however highlight similarities with the approach in \cite{manybody}.

The main result of this paper will be the following expression for the leading period de Haas - van Alphen magnetic oscillations in a class of 2+1 dimensional theories that exhibit an emergent quantum criticality at low energies:
\be\label{eq:finalresult}
\chi_\text{osc.} = - \frac{\pa^2 \Omega_\text{osc.}}{\pa B^2} = \frac{ \pi A T c k_F^4}{e B^3} \cos \frac{\pi c k_F^2}{e B}\sum_{n=0}^\infty e^{- \frac{c T}{e B}\frac{k_F^2}{\mu} \, \left(\frac{T}{\mu}\right)^{2\nu-1} F_n(\n)} \,,
\ee
\noindent where $\chi$ is the magnetic susceptibility,  $e$ is the charge of a fermionic operator, $A$ is the area of the sample, $T$ the temperature, $c$ the speed of light, $k_F$ the Fermi momentum, $B$ the applied magnetic field, $\mu$ the chemical potential and $\nu$ a critical exponent.
Our computations are in the clean limit, with no disorder. The most important of these parameters, for our purposes, is the critical exponent
$\nu$, which satisfies $0 \leq \nu \leq \half$. At $\nu = \half$ we will find $F_n(\half) = 2\pi^2 \bar h(n+\half) $, where $\bar h$ is a dimensionless constant defined below. The sum in expression (\ref{eq:finalresult}) then gives
\be\label{eq:kl}
\chi_\text{osc.} = \frac{ \pi A T c k_F^4}{2 e B^3} \, \frac{\cos \frac{\pi c k_F^2}{e B} }{\sinh \frac{\pi^2 \bar h c T}{e B}\frac{k_F^2}{\mu}} \,. \qquad  \qquad (\nu= \half)
\ee
This is essentially the textbook Lifshitz-Kosevich result \cite{abrikosov, LifKos}, as we discuss in more detail below. Our theories will be in 2+1 dimensions, although many results can likely be generalized to 3+1 dimensions.
When $\nu < \half$ the functions $F_n(\n)$, given below, are considerably more complicated. The point we wish to emphasize, however, is that at larger temperatures $T \gtrsim \frac{\mu e B}{c k_F^2}$, the decay of the amplitude as a function of $T$ is not of the simple exponential form predicted by the Lifshitz-Kosevich formula, but rather
\be\label{eq:nkl}
\chi_\text{osc.} \sim e^{- T^{2\nu}} \,. 
\ee
This is what we will mean by a generalized Lifshitz-Kosevich scaling. If we write this scaling as a temperature dependent effective quasiparticle mass in the usual Lifshitz-Kosevich formula, then
\be
m_\star \sim \frac{k_F^2}{\mu} \left(\frac{\mu}{T} \right)^{1-2 \nu } \,,
\ee
which is divergent when $T \ll \mu$ and $\nu < \half$. This is perhaps interesting in the light of the observations in \cite{suchitra, heavy}.

\begin{figure}[h]
  \begin{center}
    \includegraphics[width=5in]{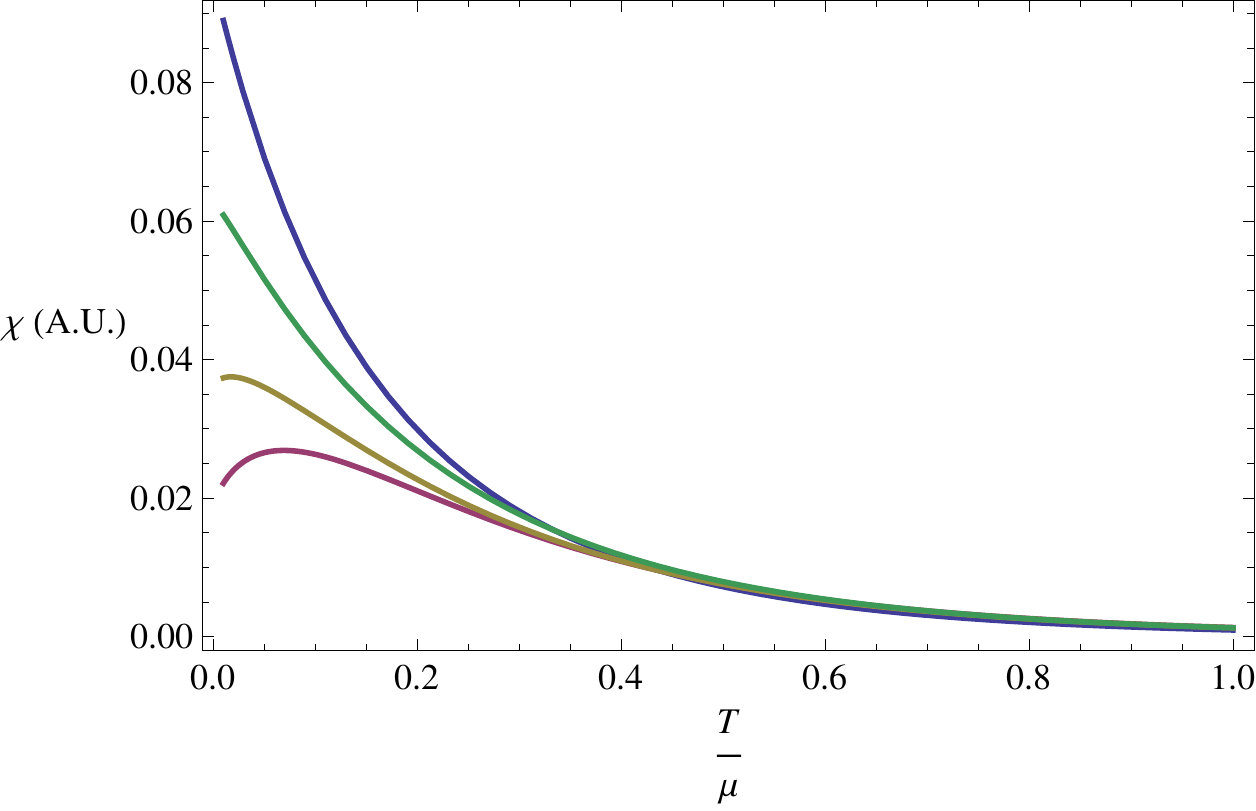}
  \end{center}
   \caption{\small Typical dependences of the amplitude of quantum oscillations on temperature. For illustration $\nu = \frac{1}{3}$, $\frac{eB}{c k_F^2}=1$, $\a=1$. Angles of $\hat h$ from top to bottom: $\varphi = \{-\varphi_0$, $-0.2 \varphi_0$, $0.51 \varphi_0$, $\varphi_0\}$ where the maximum value $\varphi_0 \equiv \pi (\half - \nu)$. The magnitude of $\hat h$ has been scaled to make the large temperature behavior coincide: $h = \{0.34, 0.39, 0.58,1\}$.}
  \label{chiplot}
  \end{figure}

While the large temperature scaling (\ref{eq:nkl}) is the most universal feature of our results, we can also plot the full amplitude (\ref{eq:finalresult}) as a function of temperature for given values of the parameters. We will introduce the various free parameters of the model below. Typical results are shown in figure \ref{chiplot}. The most interesting observation is that for a given value of the critical exponent $\nu < \half$ there is a range of possible behaviors at low temperature. While the curves can saturate, mimicking the usual Lifshitz-Kosevich behavior, it is also possible for the curve to reach zero temperature with a finite negative gradient or alternatively to exhibit a maximum before reaching zero temperature with a positive gradient. A maximum was reported experimentally in \cite{maximum}. In \cite{maximum} it was further noted that an improved fit to the data could be achieved by modifying the Kosevich-Lifshitz formula.

The theories for which (\ref{eq:finalresult}) will be shown to hold are described using the `holographic correspondence'. We will not review the methodology in detail, introductions written for the condensed matter community can be found in \cite{McGreevy:2009xe, Hartnoll:2009qx}, but rather summarise the physical properties of the theories in question.

The holographic correspondence allows a class of strongly interacting quantum field theories to be studied in a limit in which there are a large number of degrees of freedom per site. Unlike more traditional vector `large $N$' limits, the theories do not become weakly interacting in this limit, and might therefore be expected to capture aspects of interesting experimental systems that would otherwise elude theoretical control.

It was shown in \cite{Faulkner:2009wj}, following earlier work in \cite{Lee:2008xf, Liu:2009dm, Cubrovic:2009ye}, that the fermion spectral densities in these theories exhibit a broad peak with a zero temperature dispersion relation at $k \sim k_F$ of the form
\be\label{eq:dispersion}
\frac{\omega}{v_F} + h e^{i \theta} \omega^{2 \nu} = k - k_F \,, 
\ee
where $\{v_F, h, \theta, \nu,k_F\}$ are real constants. For $\nu < \half$ the non-analytic term $\w^{2 \nu}$ dominates at low frequencies, leading to non-Fermi liquid behaviour. The case $\nu = \half$ leads to the dispersion relation $\frac{\omega}{v_F} + h e^{i \theta} \omega \log \w = k - k_F$, which is precisely that of a marginal Fermi liquid \cite{marginal}. For all $\nu \leq \half$, the peak in the spectral density does not correspond to a stable quasiparticle excitation. This is because the width of the peak is always comparable to its height. Viewed as a pole in the spectral density in complex frequency space, its residue goes to zero as the pole hits the real axis at $k=k_F$ \cite{Faulkner:2009wj}. In principle we could also study $\nu > \half$, but here the linear term in (\ref{eq:dispersion}) dominates at low energies and a more conventional behaviour is expected. See however \cite{Faulkner:2009wj} for some curious properties of these cases.

Given that (\ref{eq:dispersion}) does not describe a weakly interacting (stable) quasiparticle, one can anticipate that the contribution of the fermions to thermodynamic and transport quantities will not be simply that of a free fermion with dispersion (\ref{eq:dispersion}). The correct way of computing in these systems was developed in \cite{Denef:2009yy}, with the more mathematical aspects treated in \cite{Denef:2009kn}. The essential step is to consider (\ref{eq:dispersion}) as the singular locus of the fermion spectral density $\text{Im} G_R(\w,k)$. It is easy to see that (\ref{eq:dispersion}) has two types of singularities, a pole and then a branch cut emanating from $\w = 0$. While the pole describes the na\"ive `quasiparticle', both the pole and the branch cut will give contributions to e.g. thermodynamic quantities.

This paper will be concerned with small but finite temperatures. At finite temperature, the branch cut of (\ref{eq:dispersion}) is resolved into closely spaced poles. For $T,\w \ll \mu$ one obtains \cite{Faulkner:2009wj} that the poles of $\text{Im} G_R(\w,k)$ are given by solutions to
\be\label{eq:smallw}
{\mathcal{F}}(\w_\star(k)) = 0 \,,
\ee
where
\be\label{eq:curlyF}
{\mathcal{F}}(\w) = \frac{k-k_F}{\Gamma\left(\half + \nu - \frac{i \w}{2 \pi T} - i \a \right)} -
\frac{h e^{i \theta} e^{i \pi \nu} (2 \pi T)^{2 \nu}}{\Gamma \left(\half - \nu - \frac{i \w}{2 \pi T} - i \a \right)} \,.
\ee
See e.g. figure (3) of \cite{Denef:2009yy}. The dimensionless constant $\alpha$ is related to the normalisation of the current-current correlator \cite{Faulkner:2009wj}. While complicated, this formula is largely fixed by an emergent $SL(2,\R)$ (or possibly even Virasoro) symmetry at energies $\w \ll \mu$, suggesting perhaps validity beyond the specific holographic theories considered in \cite{Faulkner:2009wj}. This emergent IR scaling symmetry is the quantum criticality referred to in the title of this paper. The only dimensionful scales in the theory are the chemical potential $\mu$, magnetic field $B$, Fermi momentum $k_F$ and temperature $T$. In (\ref{eq:curlyF}) we have assumed that $\nu < \half$ so that the linear in $\w$ term in (\ref{eq:dispersion}) can be dropped at low energies.

All of the poles given by (\ref{eq:smallw}) contribute to quantities of interest, even those that are a long way away from the real frequency axis. The key result of \cite{Denef:2009yy, Denef:2009kn} was to express the contribution of the fermions to the free energy as a sum of contributions from these poles. The formula is
\be\label{eq:formula}
\Omega = \frac{e B A T}{2 \pi c}  \sum_\ell \sum_{\w_\star(\ell)} \log \left( \frac{1}{2 \pi}
\left| \G \left(\frac{i \w_\star(\ell)}{2 \pi T} + \frac{1}{2} \right) \right|^2 \right) \,.
\ee
Anticipating our interest in magnetic fields, we have given the free energy as a sum over Landau levels rather than momenta. The first term in (\ref{eq:formula}) is the degeneracy of the Landau levels. The frequencies $\w_\star(\ell)$ are obtained from $\w_\star(k)$ in (\ref{eq:smallw}) and (\ref{eq:curlyF}) by the replacement $k^2 \to \frac{  2\ell e B}{c}$. This replacement is precise in the limit $\frac{e B}{c} \ll k_F^2$ that we will be interested in. The formula (\ref{eq:formula}) is not as exotic as it may appear; for instance, the free energy of a damped harmonic oscillator can be computed using essentially the same formula, with $\w_\star$ again given by the poles of the retarded Green's function \cite{Denef:2009yy, Denef:2009kn}. The appearance of $|\G(i x + \half)|^2$ is a generalisation of the Fermi-Dirac distribution to complex energies. If $x$ is real then $|\G(i x + \half)|^2 = \pi\,  \text{sech} \pi x$, recovering the standard expression.

Our objective is to perform the sum (\ref{eq:formula}) given (\ref{eq:curlyF}) to obtain the magnetic susceptibility for general $T \sim \frac{e B}{m_\star c} \ll \mu$. The result for the leading oscillatory part of the susceptibility is stated in (\ref{eq:finalresult}).

\section{The computation}

Our starting point is the formula for the fermionic contribution to the free energy, given in (\ref{eq:formula}) in terms of the poles (\ref{eq:smallw}) of the fermion retarded Green's function. It will be useful to consider the dimensionless quantity
\begin{equation}\label{eq:dimless}
\hat \Omega \equiv \frac{2\pi c}{e B A T} \Omega = \text{Re} \sum_\ell \sum_{x_\star(\ell)} 2 \log \Gamma\left(x_\star(\ell) + \frac{1}{2}\right) \,,
\end{equation}
where we set
\be
x = \frac{ i \w}{2 \pi T} \,.
\ee
In the formula (\ref{eq:curlyF}) defining the poles  we will furthermore set
\be\label{eq:redef}
\hat{h} \equiv \frac{h e^{i\theta + i \pi \nu} (2\pi \mu)^{2\nu}}{\pi k_F} \equiv \bar{h}(\sin \varphi + i \cos \varphi) \,,
\ee
so that $\{\hat h, \bar h, \varphi \}$ are now dimensionless. While in principle these parameters are determined by data in the UV by solving some ordinary differential equations numerically \cite{Faulkner:2009wj}, we will simply treat them as order one quantities, as we are more interested in parametrising possible low energy physics. There is a restriction on $\varphi$ that ensures that the poles are in the lower half frequency plane: $-\pi (\frac{1}{2}-\nu) < \varphi < \pi (\frac{1}{2}-\nu)$. Notice that the imaginary part of $\hat{h}$ is always positive.

Using all these expressions we can rewrite the sum over $\w_\star(\ell)$ as a contour integral. Noticing that ${\mathcal{F}}(x)$ does not have poles, just zeroes in the right half plane (corresponding to the poles $\w_\star$ of $G_R(\w,k)$ in the lower half plane) we can write
\begin{equation}\label{step1}
\hat \Omega = \text{Re} \frac{i}{ \pi} \sum_\ell \int_{-  \frac{1}{4} - i \infty}^{-  \frac{1}{4} + i \infty} dx  \log \Gamma\left( x + \frac{1}{2}\right) \frac{{\mathcal{F}}'(x)}{{\mathcal{F}}(x)} \,.
\end{equation}
The contour was chosen such that it leaves the poles of $\frac{{\mathcal{F}}'(x)}{{\mathcal{F}}(x)}$ to the right and the branch cut of $\log \Gamma\left( x + \frac{1}{2}\right)$ to the left. Implicitly we are also taking the contour to include a large semicircle in the right half plane. We will not need to evaluate the contribution from the semicircle explicitly, at a later step we will exhange the current sum over poles inside the contour for a sum of poles outside the countour (i.e. in the left hand plane).

We would like to integrate (\ref{step1}) by parts, but this is complicated by the presence of the branch cuts from the logarithmic term. However, the derivative of $\hat \Omega$ with respect to the magnetic field can be safely integrated by parts to give
\be
\hat M \equiv \frac{\pa \hat \Omega}{\pa B} =  \text{Re} \frac{1}{ i \pi} \sum_\ell  \int_{- \frac{1}{4} - i \infty}^{- \frac{1}{4} + i \infty} dx \frac{\Gamma'\left(x+\frac{1}{2}\right)}{\Gamma\left(x+\frac{1}{2}\right)}  \frac{\partial_B {\mathcal{F}}(x,B)}{{\mathcal{F}}(x,B)} \,.\label{M1}
\ee
We will be interested in considering the periodic behavior in $\frac{1}{B}$ of this expression. Therefore, it is of use to Fourier transform the Landau level variable $\ell$. We will perform a Poisson resummation to rewrite (\ref{M1}). The formula we use is
\begin{equation}
\sum_{\ell=0}^{\infty} f(\ell) =\sum_{k=-\infty}^{\infty} \int_{0^-}^{+\infty} dx\,\, f(x) e^{i 2 \pi k x} \label{prs} \,.
\end{equation}
It is straightforward to apply this formula to (\ref{M1}), with the Landau levels going over $\ell = 0,1,2,\ldots$. We obtain
\begin{equation}
\hat M = \text{Re} \frac{1}{ i \pi} \sum_{k=-\infty}^{\infty} \int_{- \frac{1}{4} - i \infty}^{- \frac{1}{4} + i \infty} dx \frac{\Gamma'\left(x+\frac{1}{2}\right)}{\Gamma\left(x+\frac{1}{2}\right)} G(x,B,k) \,,
\label{M2}
\end{equation}
where
\be
G(x,B,k) \equiv \int_0^\infty d\ell\,\, \frac{\partial_B {\mathcal{F}}(x,B,\ell)}{{\mathcal{F}}(x,B,\ell)} e^{i 2 \pi k \ell} = 
\frac{c k_F^2}{2 e B^2} \int_0^\infty \frac{du\,\, u^2\,\, e^{i 2 \pi \frac{c k_F^2}{2 e B} k u^2}}{u - \left(1+ \pi\hat h \left(\frac{T}{\mu}\right)^{2\nu} S_\nu(x) \right)}\label{G1}
\,.
\ee
in which we used the explicit form of (\ref{eq:curlyF}), changed variables to $u=\sqrt{\frac{2 e B\ell}{ ck_F^2}}$ and set
\be\label{eq:ss}
S_\nu(x) = \frac{\Gamma\left(\frac{1}{2} + \nu - i \alpha -x\right)}{\Gamma\left(\frac{1}{2} - \nu - i \alpha -x\right)} \,.
\ee 

Equations (\ref{M2}) and (\ref{G1}) appear to involve formidable sums and integrals. However, we can now neatly
separate out the oscillating and non-oscillating parts of this expression. We will deform the contour in such a way that the integral follows a steepest descent path of the exponential term. The reason this helps is that the resulting integral is manifestly non-oscillating in $1/B$. 

We therefore deform the integral in (\ref{G1}) by $u \rightarrow e^{i\frac{k}{|k|}\frac{\pi}{4}}  u$. It is crucial to realize here that the contour needs to be rotated in opposite directions in the complex plane, depending on the sign of $k$, to guarantee convergence. The only possible obstructions to this contour rotation are either a contribution at infinity, which is absent in our case as the integrands decay exponentially if the paths are rotated in the correct direction, or if a pole is crossed as the contour is deformed. The expression (\ref{G1}) makes manifest that there is such a pole at $u= 1 + \pi\hat h \left(\frac{T}{\mu}\right)^{2\nu} S_\nu(x)$. In the limit of physical interest, $T/\mu \rightarrow 0$, this pole is slightly off the real axis, for $0<\nu<\frac{1}{2}$, where our formulae are valid. 

The exact position of the pole depends on the phase of $\hat{h}$ but it is always slightly above the real axis (this can easily be checked for the allowed range of values of $\hat h$ and $x \in -\frac{1}{4} + i \mathbb{R}$). The upshot is that for negative $k$ we can rotate the contour and get
\begin{equation}
G(x,B,-|k|) = \frac{c k_F^2}{2 e B^2} e^{-i\frac{\pi}{2}}  \int_0^\infty du\,\, \frac{u^2\,\, e^{- 2 \pi \frac{c k_F^2}{2 e B^2}|k| u^2}}{u - e^{i\frac{\pi}{4}}  \left(1+\pi\hat h \left(\frac{T}{\mu}\right)^{2 \nu} S_\nu(x)\right)} \,. \label{Gn} 
\end{equation}
This contribution is strictly non-oscillating\footnote{A more formal way to state this fact is that upon Fourier transforming in $\frac{1}{B}$, the transform has poles only at imaginary frequencies.} in $\frac{1}{|B|}$.
Deforming the contour for positive $k$ we pick up a contribution from the pole. Calculating the appropriate residue yields
\bea
G(x,B,|k|) &=& G_\text{non-osc.}(x,B,|k|) + G_\text{osc.}(x,B,|k|) \label{Gp} \\
&=& G_\text{non-osc.}(x,B,|k|)  +  \frac{\pi i c k_F^2}{e B^2}   \left(1+\pi\hat h {\textstyle \left(\frac{T}{\mu}\right)^{2 \nu} } S_\nu(x)\right)^2 e^{i 2 \pi \frac{c k_F^2}{2 e B} |k|   \left(1+\pi\hat h \left(\frac{T}{\mu}\right)^{2 \nu} S_\nu(x)\right)^2} \nonumber \,.
\eea
The first term is non-oscillating and is the same as (\ref{Gn}) with various factors of $e^{i \pi/4} \to e^{-i\pi/4}$.
We are therefore left with the following oscillating contribution
\begin{equation}\label{eq:osc}
G_\text{osc.}(x,B,k) = \Theta(k) G_\text{osc.}(x,B,|k|) \,.
\end{equation}
where $\Theta(k)= 1$ for $k>0$ and $\Theta(k)= 0$ for $k<0$ . The $k=0$ term is also non-oscillating and does not concern us.
We have thus performed the first of our integrals, insofar as obtaining the oscillating term is concerned.

The next integral to address is the $x$ integral in (\ref{M2}). We will convert this integral into a sum over residues that are {\it outside} the original region of integration. That is, to the left of the imaginary axis. Doing this allows us to represent the integral as a sum of the residues of the poles of $ \frac{\Gamma'(x+\frac{1}{2})}{\Gamma(x+\frac{1}{2})} $. These are located at $-\frac{1}{2}-n$ with $n=0,1,2,3,\ldots$ and have minus unit residue. Combining this operation with the result (\ref{eq:osc}),
our expression (\ref{M2}) becomes
\begin{eqnarray}
\hat M_\text{osc.} = \text{Re} \frac{2 \pi c k_F^2}{i e B^2} \sum_{k=1}^\infty \sum_{n=0}^\infty \left(1+\pi\hat h {\textstyle \left(\frac{T}{\mu}\right)^{2 \nu} } S_\nu(-\half-n)\right)^2  e^{i 2 \pi \frac{c k_F^2}{2 e B} k  \left(1+\pi\hat h \left(\frac{T}{\mu}\right)^{2 \nu} S_\nu(-\frac{1}{2}-n)\right)^2}\,.\label{M4}
\end{eqnarray}
It is clear at this point that we have obtained sums over terms that both oscillate and decay in $1/B$. We can now take the physical $T/\mu \to 0$ limit keeping only leading terms determining the oscillations and exponential decay. The result is
\begin{eqnarray}
\hat M_\text{osc.} = \frac{2 \pi c k_F^2}{e B^2} \sum_{k=1}^\infty \sin {\frac{\pi c k_F^2 k}{e B} } \sum_{n=0}^\infty  e^{- 2 \pi^2 \frac{c k_F^2}{e B}  \left(\frac{T}{\mu}\right)^{2\nu}  k\, \textrm{Im} \, \hat h \,S_\nu\left(-\half - n\right)} \,.  \label{M6}
\end{eqnarray}

This last formula is essentially the result. To compute the magnetic susceptibility $\chi$ we have to reinsert the factors that relate
$\Omega$ to $\hat \Omega$ in (\ref{eq:dimless}). Thus
\begin{equation}\label{chiM}
\chi = - \frac{\pa^2 \Omega}{\pa B^2}  =  - \frac{e A T}{\pi c} \hat M - \frac{e B A T}{2\pi c}  \frac{\partial \hat M}{\partial B} \,.
\end{equation}
For situations of physical interest we have $\frac{eB}{c} \ll k_F^2$ and therefore the leading result comes from the second term by acting with the derivative on the sine in (\ref{M6}). Focusing on the leading period, the $k=1$ term, this gives our main result, that we already quoted in equation (\ref{eq:finalresult}),
with
\be
F_n(\nu) = 2\pi^2  \text{Im} \, \hat h \, S_\nu\left(-\half - n\right) \,.
\ee

We also already noted in the introduction that the case $\nu = \half$ is special. This is because the ratio of gamma functions in (\ref{eq:ss}) simplifies in this case to give $F_n(\half) = 2\pi^2  \bar h (n+\half)$. The sum over $n$ can then be done explicitly, to yield a result of the standard Lifshitz-Kosevich form (\ref{eq:kl}).
 
In general, we cannot perform the sum over $n$ in closed form. However, it is simple to check numerically that for all allowed values of the parameters, $F_n(\nu)$ is positive and monotonically increasing in $n$. Therefore at the high temperatures of primary interest we can keep only the first term in the sum in (\ref{M6}) or (\ref{eq:finalresult}) given by $n=0$. This observation also implies that the $k=1$ term kept in (\ref{eq:finalresult}) has an exponentially larger amplitude than the other terms in this regime.
Thus we obtain, for general $\nu < \half$, the non-Lifshitz-Kosevich scaling that we quoted in (\ref{eq:nkl}).

\section{A general formula for quantum oscillations}

We will now rederive the result (\ref{eq:finalresult}) via a slick argument. The argument is quite general and we anticipate future applications.
%We will show that textbook results for free fermions can be rederived in a few lines.
The method used is a generalisation of that in \cite{Denef:2009kn}\footnote{Frederik Denef has been privately advocating this type of generalisation for some months.} and we will be brief in presentation.

The statement is that for any fermionic system satisying assumptions to be given shortly
\be\label{eq:coolexpression}
\Omega_\text{osc.} = \frac{e B A T}{\pi c}  \, \text{Re} \, \sum_{n=0}^\infty \sum_{k=1}^\infty \frac{1}{k} \, e^{i 2\pi k \ell_\star (n)}
\,, 
\ee
where the $\ell_\star(n)$ are defined as the solutions to
\be\label{eq:gensol}
{\mathcal{F}}(\w_n,\ell_\star(n)) = 0 \,.
\ee
Here ${\mathcal{F}}(\w,\ell)=0$ defines the singular locus of the fermion retarded Green's function in a magnetic field, $G_R(\w,\ell)$. The fermionic Matsubara frequencies are $\w_n = 2 \pi i T \left(n + \half \right)$. We assume for simplicity that there is a unique $\ell_\star(n)$, but it is simple to relax this assumption. It is clear that using (\ref{eq:curlyF}) with $k^2 = \frac{2 \ell e B}{c}$, solving for $\ell_\star(n)$ as in (\ref{eq:gensol}) and plugging into (\ref{eq:coolexpression}) immediately reproduces our previous result (\ref{M4}).

The class of theories to which the formula (\ref{eq:coolexpression}) will most directly apply are those where the fermionic partition function can be expressed as the determinant of an operator $\ocal$ in a thermal Euclidean space. This certainly applies to free theories and to theories with classical holographic duals. In the latter case the determinant is in one extra curved spacetime dimension, but this does not make a difference to the argument. We assume that in a background magnetic field, the eigenvalues of the operator can be labelled by the quantum numbers $\w_n$ and $\ell$ as well as any others. The type of reasoning in \cite{Denef:2009kn} is quickly seen to imply that we must have, up to UV contributions that can be dealt with systematically but which will not contribute to oscillations,
\be\label{eq:neat}
\Omega = - T \tr \log \ocal = - \frac{e B A T}{\pi c} \text{Re} \sum_{\w_n \geq 0} \sum_\ell \log (\ell - \ell_\star(n)) \,.
\ee
The logic that leads to this expression is to separate the eigenvalues of $\ocal$ according to $\w_n$ and $\ell$.
The contribution from positive and negative $\w_n$ to the determinant are complex conjugates of each other \cite{Denef:2009yy, Denef:2009kn} so we concentrate on the positive Matsubara frequencies.
For a fixed $\w_n$ we can deform the operator by letting $\ell \to \ell + \gamma$ and then match the zeros of the determinant of $\ocal_{n,\gamma}$ as a function of $\g$. Zeros arise whenever $\ocal_{n,\gamma}$ has a zero mode. This in turn occurs whenever 
the Euclidean Green's function has a pole at $\w = \w_n$, which we define to occur at $\ell + \gamma \equiv \ell_\star(n)$.
The retarded Green's function is the analytic continuation of the Euclidean Green's function from the upper half frequency plane, thus connecting with our definition of ${\mathcal{F}}(\w,\ell)$ appearing in (\ref{eq:gensol}). Writing $\det \ocal_{n,\gamma} \sim \prod_\ell (\ell + \gamma - \ell_\star(n))$ and setting $\gamma = 0$ gives (\ref{eq:neat}).

Poisson resumming (\ref{eq:neat}) using (\ref{prs}) and picking out the oscillatory part of the Fourier transform by rotating the contour in different directions for negative and positive $k$, in a similar way to how we did previously, then directly leads to (\ref{eq:coolexpression}). Only the rotation at positive $k$ leads to a singularity contribution giving the oscillating term.

We now see that the formula (\ref{eq:coolexpression}) reproduces known expressions for free fermions. The non relativistic, spinless electron (the effect of spin is simply to multiply the answer by two in the limit $\frac{eB}{c} \ll k_F^2$) has
\begin{equation}\label{eq:Ffree}
{\mathcal{F}}_\text{non-rel.}(\w,\ell) =  \frac{B e}{m c} \ell - \mu - \w \,.
\end{equation}
It is trivial to solve for $\ell_\star(n)$ defined via (\ref{eq:gensol}). Plugging into (\ref{eq:coolexpression}), differentiating twice and performing the geometric series sum over $n$ as previously leads to
\begin{equation}
\chi_\text{osc.} = \frac{2 \pi A T \mu^2 \, m^2 c }{B^3 e } \sum_{k=1}^\infty  \frac{k \, \cos \left( 2\pi k \frac{\mu \, m c}{B e}\right)}{\sinh\left(2 \pi^2 k \frac{ T \, m c}{B e} \right)}  \,. \label{chiNR}
\end{equation}
This is literally the Lifshitz-Kosevich formula \cite{abrikosov} in 2+1 dimensions, which we have derived rather painlessly. The fact that ${\mathcal{F}}$ is linear in $\ell$ in (\ref{eq:Ffree}) makes the steps leading to (\ref{eq:neat}) trivial in this case, there is no rewriting involved.

We can treat the spinless relativistic fermion similarly. In this case
\begin{equation}
{\mathcal{F}}_\text{rel.}(\w,\ell) = m^2 c^4+ 2 B e \ell c - (\mu + \w)^2 \,.
\end{equation}
It is again immediate to solve for $\ell$. Use of  (\ref{eq:coolexpression}), the limit $T \ll \mu$, differentiation and summing a geometric series gives 
\be
\chi_\text{osc.} = \frac{ \pi A T c k_F^4}{2 B^3 e} \sum_{k=1}^\infty  k \,\frac{\cos  \left( \pi k \frac{c k_F^2}{B e}\right)}{ \sinh \left(\pi^2 k \frac{ T \, \mu}{B e c}\right)}
\ee
We used the relation $k_F^2 c^2 = m^2 c^4 - \mu^4$.
In the massless limit (or $\mu$ much larger than $m c^2$) this expression recovers our result (\ref{eq:kl}) for the `marginal' non-Fermi liquid at $\nu=\frac{1}{2}$ if we choose $\bar{h}=1$.

The expression (\ref{eq:coolexpression}) is essentially the same as a general expression appearing in \cite{manybody}. In \cite{manybody} the effects of interactions are incorporated into a renormalised self energy for quasiparticles whose one loop contribution to the susceptibility is then computed. This is a controlled approximation if there are well defined quasiparticles so that higher order corrections that cannot be absorbed into the self energy are negligible. In the holographic theories studied here the self energy due to strong interactions is captured by the propagation of the fermions on a nontrivial background spacetime, leading to the singular locus (\ref{eq:curlyF}). Interactions between these fermions are suppressed by the `large N' limit in which the holographic computations are performed. Therefore, holography provides a controlled setting in which the self energy can be strongly renormalised to the extent that there are not well defined quasiparticles and yet quantities such as the susceptibility can be computed with a determinant formula like (\ref{eq:coolexpression}).

 \section{Magnitude of oscillations and the Fermi surface}

We need to check that our result could in principle be measured. For that purpose we compare the order of magnitude of the amplitude of the oscillating part to the non-oscillating part. We will pursue this calculation at low temperatures, where the oscillating signal is strongest. In this limit, we will see that  the oscillating susceptibility strongly dominates over the non-oscillating part in the regime of interest $\frac{eB}{c} \ll k_F^2$ for $\frac{1}{6}<\nu<\frac{1}{2}$. This dominance is, of course, not a strict requirement for experimental detection. 
We first estimate the oscillating magnetization. At low temperatures all terms in the sum in (\ref{eq:finalresult}) are important. In fact, the infinite tail of this sum dominates. Therefore, we can replace $S_\nu(-\frac{1}{2} - n)$ with its $n \rightarrow \infty$ limit, $S_\nu(-\frac{1}{2} -n) \rightarrow  n^{2 \nu}$. Because the quantity appearing in the sum is $\frac{T}{\mu}n$, we can replace the sum in $n$ with an integral at leading order in $\frac{T}{\mu}$. Therefore the magnitude of (\ref{eq:finalresult}) becomes
\begin{equation}\label{zTno}
\frac{\left| \chi^{T \sim 0}_{\textrm{osc}} \right|}{A} \sim \frac{c k_F^4 T}{e B^3} \int dn \, e^{- \frac{c \k_F^2}{e B} \, (\frac{T}{\mu})^{2\nu} n^{2\nu}} \sim \frac{e^2 \mu}{c^2 k_F^2}  \times \left(\frac{e B}{c  k_F^2}\right)^{\frac{1}{2\nu}-3} \,.
\end{equation} 

It is interesting to rederive this last result from a different perspective that makes transparent the role of a Fermi surface.
At low temperatures the susceptibility is most naturally written as a sum over $\ell$, without Poisson ressumming. 
We can start from the expression (\ref{M1}) for $\hat M$ and calculate $\chi$ by use of (\ref{chiM}). As before, we can change the $x$ integral to a sum over poles labeled by $n$. Once again, at zero temperature the tail of this sum dominates and we can substitute $\sum_n \rightarrow \int dn$ and $S_\nu(-\frac{1}{2} -n) \rightarrow  n^{2 \nu}$. The resulting integral can be performed analytically to leave a sum over $\ell$ that is similar to the expressions obtained in \cite{Denef:2009yy}. This sum has a nonanalyticity at $\ell=\frac{c k_F^2}{2 eB}$. Expanding the
susceptibility at small $\frac{eB}{c k_F^2}$ using a generalized version of the Euler-Maclaurin formula \cite{Lyness}, the sum in $\ell$ becomes an integral plus contributions at the edges. The edge near the Fermi surface is responsible for the leading effect we are interested in. Explicitly
\be\label{eml}
\frac{\left| \chi^{T \sim 0} \right|}{A}  \sim \sum_\ell \, \frac{\mu e  }{c B}\, g\left(\frac{2 e B \ell}{c k_F^2}\right) \left(1 - \sqrt{ \frac{2 e B \ell}{c k_F^2}} \right)^{-2+1/2\nu} \sim \textrm{Analytic}(B) +  \frac{e^2 \mu}{c^2 k_F^2}  \left(\frac{e B}{c k_F^2}\right)^{\frac{1}{2\nu}-3}  
\ee
\noindent where $g(\cdot)$ is a dimensionless function that is regular at 1. The analytic terms give a generic expansion, with the constant term representing, for instance, Landau diamagnetism. This piece includes contributions that have not been captured by the poles in (\ref{eq:curlyF}), as this formula has zoomed in on the low energy states near the Fermi surface.
The second term is the leading contribution coming from the Fermi surface and agrees with the previous computation (\ref{zTno}).
From (\ref{eml}) we can see that the oscillating term strongly dominates the susceptibility for $\frac{1}{6}<\nu<\frac{1}{2}$.

Finally, we can check that the scaling (\ref{eq:nkl}) is potentially observable in an experimentally interesting regime without being exponentially suppressed by temperature. Setting all dimensionless parameters except for $\nu$ to be order unity, we can estimate the magnitude of the oscillations. Taking $\mu$ to be of order eV, $T$ to be of order Kelvin and reinserting fundamental constants the exponent in our final result (\ref{eq:finalresult}) is of order
\be\label{eq:units}
\frac{c k_F^2}{\hbar \, e B} \left(\frac{k_B T}{\mu}\right)^{2 \nu} \sim  \frac{F_B}{B}\times \left(10^{-4}\right)^{2\nu} \,,
\ee
where $F_B$ is the frequency of the oscillations measured in Tesla. In measurements on the underdoped cuprates, for instance, $F_B/B \sim 10$ \cite{Nat1}, and so the exponent is not too large for a wide range of values of $\nu$.

%The first term required us to introduce a velocity which we have written as $c = \a \, \text{ms}^{-1}$. We also set $q=e$, the electron charge. If we take a typical electronic velocity scale, then $\a^2 \sim 10^{11}$, so the first term in (\ref{eq:units}) leads to a number slightly smaller than order one multiplied by $tm/b$.

%From the second term in (\ref{eq:units}) it becomes clear that having $\nu < \half$ decreases the amplitude of the oscillations, making the negative exponent in (\ref{eq:finalresult}) bigger, relative to the Lifshitz-Kosevich case. If $t$ and $m$ are both order one, as is typical, then this is potentially worrisome. However, in many recent experiments, $b$ is large, of order 50 or higher in e.g. \cite{suchitra}. This suffices to bring the exponent back to being an order one number.

%The upshot is that at sufficiently large, but experimentally accessible, magnetic fields (or, alternatively, lower temperature) the non-Lifshitz-Kosevich scaling of (\ref{eq:finalresult}) should be distinguishable. The closer $\nu$ is to $\half$, the larger the signal.

\section{Discussion}

Using the holographic correspondence we have obtained the amplitude of quantum oscillations in a family of strongly interacting quantum critical theories. Our expression (\ref{eq:finalresult}) provides a theoretical template for possible violation of
Lifshitz-Kosevich scaling of the amplitude with temperature due to strong interactions. We also found that at the marginal value of the critical exponent $\nu = \half$, the Lifshitz-Kosevich result (\ref{eq:kl}) survives the interactions. Our results are perhaps the most concrete yet to emerge from applications of holography to condensed matter physics. The scalings we have described could conceivably be found in systems of current experimental interest.

It will be important to generalise out computations to include disorder and to see to what extent the textbook Dingle scaling is modified. The dynamics of holographic theories with disorder has barely been studied \cite{Hartnoll:2008hs}. Furthermore, while
the singular loci (\ref{eq:curlyF}) for the Green's function is the simplest following from the holographic correspondence \cite{Faulkner:2009wj}, it is likely not unique. As new finite density dual geometries become available, it will be of interest to see to what extent our result (\ref{eq:finalresult}) is modified.

\section*{Acknowledgements} It is a pleasure to acknowledge discussions with Frederik Denef and Subir Sachdev. D.M.H. would like to dedicate this paper to the memory of his uncle, Daniel Grempel, from whom he first heard about non Fermi liquids.

The research of S.A.H. is partially supported by DOE grant DE-FG02-91ER40654 and the FQXi foundation. D.M.H. would like to thank the Center for the Fundamental Laws of Nature at Harvard University for support.

\end{document}